\documentclass[twocolumn]{aastex7}
\usepackage{natbib}
\usepackage{needspace}
\usepackage{booktabs}
\usepackage{amsmath}
\usepackage{tabularx} 
\usepackage{textcomp}
\usepackage{comment}
\usepackage{subcaption}

\newcommand\vsini{$v\sin{i}$}
\newcommand\Msun{$M_{\odot}$} %
\newcommand\Rsun{$R_{\odot}$}
\newcommand\MJ{$M_{\mathrm{J}}$}
\newcommand\RJ{$R_{\mathrm{J}}$}

\newcommand\exofast{$\texttt{EXOFASTv2}$}

\newcommand{\ewli}{$93.83 \pm 3.58$\,m\AA{}}
\newcommand{\ewlife}{$75.39 \pm 3.58$\,m\AA{}}
\newcommand{\ali}{$2.49 \pm 0.12$\,dex}
\newcommand{\pmass}{$22$\,$M_{ \rm J}$}

\newcommand{\percentile}{98.4th}

\newcommand{\populationsize}{61}

\shortauthors{Kotten et al.}
\begin{document}

\title{
Lithium Enrichment in a Subgiant Star with a Brown Dwarf Companion: \\ A Planetary Engulfment Candidate}

\author[0009-0008-5864-9415]{Brooke Kotten}
\email[show]{bkotten@umich.edu}
\affiliation{Department of Astronomy, University of Michigan, Ann Arbor, MI 48109, USA}
\affiliation{Department of Astronomy,  University of Wisconsin--Madison, 475 N.~Charter St., Madison, WI 53706, USA}

\author[orcid=0000-0001-7493-7419,sname='Soares-Furtado']{Melinda Soares-Furtado}
\affiliation{Department of Astronomy,  University of Wisconsin--Madison, 475 N.~Charter St., Madison, WI 53706, USA}
\affiliation{Department of Physics, 2320 Chamberlin Hall, University of Wisconsin-Madison, 1150 University Avenue Madison, WI 53706}
\affiliation{Department of Physics \& Kavli Institute for Astrophysics and Space Research, Massachusetts Institute of Technology, Cambridge, MA 02139, USA}
\email{mmsoares@wisc.edu}  

\author[0000-0003-0381-1039]{Ricardo Yarza}
\altaffiliation{NASA FINESST Fellow}
\altaffiliation{Frontera Computational Science Fellow}
\affiliation{Department of Astronomy and Astrophysics, University of California, Santa Cruz, CA 95064, USA}
\affiliation{Texas Advanced Computing Center, University of Texas, Austin, TX 78759, USA}
\email{rcastroy@ucsc.edu}  

\author[0000-0002-6478-0611]{Andrew C.~Nine}
\affiliation{Department of Physics, University of Wisconsin-Whitewater, 771 W.~Starin Road, Whitewater, WI 53190, USA}
\email{ninea@uww.edu}

\author[0000-0002-4952-9007]{Seth A. Jacobson} 
\affiliation{Department of Earth and Environmental Sciences, Michigan State University, East Lansing, MI 48824, USA}
\email{seth@msu.edu}

\author[0000-0002-0701-4005]{Noah Vowell}
\affiliation{Center for Data Intensive and Time Domain Astronomy, Department of Physics and Astronomy, Michigan State University, East Lansing, MI 48824, USA}
\email{vowellno@msu.edu}

\author[0009-0001-2121-6666]{Olivia Maynard} 
\affiliation{Department of Computer Science and Engineering, Ohio State University, Columbus, OH 43210, USA}
\email{maynard.341@buckeyemail.osu.edu}

\author[0000-0001-6637-5401]{Allyson Bieryla}
\affiliation{Center for Astrophysics ${\rm \mid}$ Harvard {\rm \&} Smithsonian, 60 Garden Street, Cambridge, MA 02138, USA}
\email{abieryla@cfa.harvard.edu}

\author[0000-0001-7246-5438]{Andrew Vanderburg}
\affiliation{Center for Astrophysics ${\rm \mid}$ Harvard {\rm \&} Smithsonian, 60 Garden Street, Cambridge, MA 02138, USA}
\email{andrew.vanderburg.astro@gmail.com}

\author[0000-0002-7382-0160]{Jack Schulte}
\affiliation{Center for Data Intensive and Time Domain Astronomy, Department of Physics and Astronomy, Michigan State University, East Lansing, MI 48824, USA}
\email{jschulte@msu.edu}

\author[0000-0002-9052-382X]{Claudia Aguilera--G\'omez}
\affiliation{Instituto de Astrof\'isica, Pontificia Universidad Cat\'olica de Chile, Av.~Vicu\~na Mackenna 4860, 782-0436 Macul, Santiago, Chile }
\email{craguile@uc.cl}

\author[0000-0003-2558-3102]{Enrico Ramirez-Ruiz}
\affiliation{Department of Astronomy and Astrophysics, University of California, Santa Cruz, CA 95064, USA}
\email{enrico@ucolick.org}

\author[0000-0001-8812-0565]{Joseph E.~Rodriguez} 
\affiliation{Center for Data Intensive and Time Domain Astronomy, Department of Physics and Astronomy, Michigan State University, East Lansing, MI 48824, USA}
\email{jrod@msu.edu}

\author[0000-0001-9911-7388]{David W.~Latham}
\affiliation{Center for Astrophysics ${\rm \mid}$ Harvard {\rm \&} Smithsonian, 60 Garden Street, Cambridge, MA 02138, USA}
\email{dlatham@cfa.harvard.edu}

\begin{abstract}
Theoretical models predict that subgiants within a narrow mass regime can retain detectable lithium enrichment signatures from planetary engulfment. 
We test this prediction using TOI-5882, selected because it occupies this favorable subgiant parameter space and hosts a massive brown dwarf (\pmass{}, $P=7.1$\,d) companion capable of dynamically perturbing inner planets.
We investigate whether: (1) TOI-5882 exhibits lithium enhancement among similar subgiants, (2) planetary material would be deposited in the convective zone, and (3) the required engulfed mass lies within a plausible range for planetary engulfment.
Using spectra from the Tillinghast Reflector Echelle Spectrograph, we measured a Li I equivalent width of \ewlife{} and an abundance of A(Li)=\ali{}.
Comparing these values to a control sample of \populationsize{} subgiants from the GALactic Archaeology with HERMES (GALAH) DR4 survey, we find that TOI-5882 ranks in the \percentile{} percentile in both metrics, confirming significant lithium enrichment.
We evaluate the engulfment scenario by modeling convective zone deposition and estimating the mass required to reproduce the observed enhancement relative to the control sample.
We perform an estimate of the engulfed planetary mass incorporating CI chondritic Li abundances, as planets formed via core accretion are enriched in heavy elements and lithium partitions with these metals. 
This yields a required engulfed mass of $9$--$95\,M_\oplus$---an order of magnitude lower than the $5.6\,M_{\rm J}$ implied by proto-solar assumptions.
TOI-5882's lithium excess can plausibly result from the ingestion of a super-Earth to Neptune-mass planet, motivating further studies to test this scenario.
\end{abstract}


\section{Introduction}
\label{sec:intro}
Lithium is a fragile element in stellar interiors, yet it serves as a powerful diagnostic in stellar astrophysics. In young, low-mass stars, lithium is preserved in the outer layers, making its surface abundance a well-established indicator of youth \citep[e.g.,][]{Soderblom2010, Jeffries2014}. As stars evolve off the main sequence, their convective envelopes deepen and mix surface material with hotter interior layers where lithium is rapidly destroyed \citep[e.g.,][]{Iben1967}. As a result, lithium is depleted as the star ages, and significant surface lithium in these evolved stars is unexpected.

A small fraction of evolved stars defy this trend, exhibiting anomalously high lithium abundances \citep[e.g.,][]{Monaco2011,Aguilera_Gomez_2016, Zhou2022,Tsantaki2023}. These rare cases offer important clues about processes capable of replenishing or preserving lithium during post-main-sequence evolution. 
Critically, \citet{Soares-Furtado2021} demonstrated that subgiant stars occupy a uniquely favorable evolutionary window for detecting lithium enrichment from planetary engulfment. 
During this phase, the convective zone is massive enough to disassociate the inspiraling planet, yet shallow enough that the convective base temperature is below the lithium-burning threshold so the enrichment signature is not significantly diluted. 
Stars within this narrow evolutionary window provide testable predictions about the location in the star where planetary engulfment signatures should be observable.

We identified TOI-5882 (TIC\,232941965) as a compelling candidate to test these predictions based on two key criteria: (1) it occupies the theoretically favorable evolutionary window (that would show Li enrichment) as an early subgiant with a mass of $1.334^{+0.055}_{-0.065}$\,$M_\odot$ \citep{2025arXiv250109795V}, and (2) it hosts a massive brown dwarf companion capable of driving dynamical instability in an inner planetary system. With a mass of $22.01^{+0.61}_{-0.72}$\,\MJ{} and radius of $1.023^{+0.045}_{-0.038}$\,\RJ{}, the brown dwarf follows a close-in ($P = 7.1$\,d), mildly eccentric ($e = 0.0339 \pm 0.0041$) orbit \citep{2025arXiv250109795V}.
The companion was  originally identified in TESS \citep{Ricker2015} photometry and confirmed through spectroscopic follow-up by \citet{2025arXiv250109795V}.

TOI-5882\,b represents one of the most extreme companion architectures among subgiants: it is one of only ${\sim}40$ known transiting brown dwarfs with periods shorter than 10 days, and one of just four found orbiting a subgiant host with $\log(g) \leq 4$\,dex.
The architecture of this system raises compelling questions about its dynamical history.
This configuration provides a compelling dynamical pathway for planetary engulfment. Massive companions on tight orbits efficiently excite eccentricities in inner planetary systems through secular perturbations and resonances \citep[e.g.,][]{2005ApJ...627.1001T, Mustill2015}, driving planets onto star-grazing trajectories that lead to tidal inspiral and engulfment.

Beyond its favorable architecture, TOI-5882 satisfies additional selection criteria that make it an ideal test case. Ground-based follow-up revealed no signs of youth---no infrared excess, H-$\alpha$ emission, or rapid rotation \citep{2025arXiv250109795V}---ruling out primordial lithium retention as a trivial explanation for any observed enhancement. Additionally, high-quality spectra were available for this target, enabling precise lithium abundance measurements with archival data. 
The combination of theoretically favorable evolutionary state, extreme companion architecture capable of driving engulfment, absence of youth indicators, and high-quality spectroscopic data collectively positioned TOI-5882 as a compelling candidate for testing the planetary engulfment hypothesis.

Before attributing any observed lithium enhancement to planetary engulfment, we must evaluate alternative enrichment mechanisms.
Internal lithium production via the Cameron–Fowler mechanism involves synthesizing $^7$Be in hot interior layers and rapidly transporting it to cooler outer regions, where it decays into $^7$Li \citep{Cameron1971}. However, this process has been shown to operates effectively at late evolutionary stages (i.e., after the star evolves beyond the luminosity bump on the red giant branch, when extra mixing mechanisms are triggered; \citealt{Gilroy1991, Gratton2000}).
In early subgiants like TOI-5882, such deep mixing has not commenced, ruling out this internal mechanism.

Among the external enrichment scenarios, planetary engulfment provides the most direct explanation for lithium enhancement during the subgiant phase. The accretion of lithium-rich material from an engulfed planet can temporarily elevate the surface lithium abundance before dilution by the growing convective envelope \citep[e.g.,][]{Siess1999, Sandquist2002, Carlberg2012, Aguilera_Gomez_2016, Soares-Furtado2021, Zhou2022}. Other proposed external enrichment mechanisms, such as cosmic ray spallation \citep{Reeves1970} or accretion of classical nova ejecta \citep[e.g.,][]{1975A&A....42...55A, 1978ApJ...222..600S, 2015Natur.518..381T, Molaro2016}, are inconsistent with observed $^7$Li/$^6$Li isotopic ratios \citep[e.g.,][]{Fields1999, 1999ApJ...523..654R, 2002PhDT.........9K}, or would be expected to produce additional chemical anomalies, including altered C/N and O/N ratios and enhancements in beryllium, boron, neon, and aluminum \citep[e.g.,][]{1998PASP..110....3G, 2010AJ....140.1347H, Kemp2024}.

In this work, we test the planetary engulfment hypothesis for TOI-5882 through three complementary analyses. First, we measure TOI-5882's lithium abundance and compare it to a control sample of similar subgiants to determine whether it exhibits the predicted enhancement (Sections~\ref{sec:data}--\ref{sec:methods}). 
We note that while our analysis is differential in nature, the analysis is subject to residual systematics, including inter-pipeline parameter offsets, atomic diffusion, depletion history, and rotation-driven abundance effects.
We find that any residual inter-pipeline systematics would need to be comparable to the spread of the control sample and systematically aligned with lithium to alter our conclusions.
We model the stellar structure to assess whether planetary material can be deposited in the convective zone (Section~\ref{sec:discussion}). Third, we estimate the engulfed mass required to produce the observed lithium excess, accounting for realistic planetary compositions (Section~\ref{sec:discussion}). We summarize our findings in Section~\ref{sec:summary}.

\section{Data}\label{sec:data}
\subsection{TRES Spectroscopy}\label{subsec:tres}
\setcounter{footnote}{0}
To measure the lithium equivalent width (Li\,I EW) and abundance of TOI-5882, our team employed optical spectroscopy from the Tillinghast Reflector Echelle Spectrograph (TRES, \citealt{gaborthesis}), which is mounted on the 1.5-m Tillinghast Telescope at the Fred Lawrence Whipple Observatory in Amado, Arizona. 
These data cover an optical range of 3860-9100\,\AA{} with the resolving power of $R \approx 44,000$. 

The spectra were obtained as part of the TESS Follow-up Observing Program \citep[TFOP;][]{Collins:2018}, which consists of science working groups conducting follow-up imaging, reconnaissance spectroscopy, and precision Doppler spectroscopy to characterize TESS Objects of Interest (TOIs). TOI-5882 was flagged for follow-up observations to confirm the presence of a transiting companion.
The TRES spectra are available to the public and can be downloaded from the Exoplanet Follow-up Observing Program (ExoFOP) platform \citep{exofop}. 
Twelve TRES spectra were obtained between 2022 November 16 and December 12 with an average signal-to-noise ratio of 31 per resolution element of $6.7$\,km\,s$^{-1}$ in the peak continuum of the Mg\,b order near $5187$\,\AA. 
The details of the observations of TOI-5882 used in this analysis are listed in Table~\ref{tab:data}. 

\begin{deluxetable}{lccc}[tbh!]
\centering
\tablewidth{0pt}
\tablehead{
    \colhead{Instrument} & 
    \colhead{Date} & 
    \colhead{Exp.~Time} & 
    \colhead{SNRe} \\
    \colhead{} & 
    \colhead{(BJD)} & 
    \colhead{(s)} & 
    \colhead{}
}
\startdata
TRES & 2459899.624002 & 600 & 23.0 \\
TRES & 2459902.697074 & 600 & 31.9 \\
TRES & 2459903.669095 & 2400 & 35.3 \\
TRES & 2459905.633562 & 2100 & 25.0 \\
TRES & 2459906.706138 & 2140 & 39.0 \\
TRES & 2459912.630849 & 1350 & 29.8 \\
TRES & 2459914.568654 & 750 & 37.2 \\
TRES & 2459915.602552 & 1500 & 28.9 \\
TRES & 2459915.629738 & 1050 & 32.5 \\
TRES & 2459922.593296 & 1300 & 30.2 \\
TRES & 2459923.574595 & 900 & 32.7 \\
TRES & 2459925.606775 & 950 & 26.7 \\
\enddata
\caption{Follow-Up Spectroscopy of TOI-5882.}\label{tab:data}
\end{deluxetable}

The TRES data were blaze-corrected, as well as wavelength-corrected to account for the barycentric and radial velocity shifts present in the data. 
The spectra were then co-added to maximize the signal-to-noise ratio (SNR), achieving an SNR of 58 per pixel within the optical order of interest, as estimated using  \texttt{specutils.analysis.snr\_derived} \citep{specutils,nicholas_earl_2024_10681408}. For reference, TRES' resolution is 6.7 km\,$\text{s}^{-1}$ per resolution element.

Measurements of surface gravity ($\log{g}$), effective temperature ($T_{\mathrm{eff}}$), and metallicity ([M/H]) were obtained for TOI-5882 using the TRES Stellar Parameter Classification (SPC) pipeline \citep{2021tsc2.confE.124B, buchhave2012yCatp038048601B}. 
We allowed stellar parameters to float and opted not to implement the use of stellar isochrones, which would bias measurements toward values reflective of main-sequence stars. The results of our calculation are provided in Table~\ref{tab:5882params}, along with identifying information for TOI-5882. The table also provides stellar properties and the corresponding lithium equivalent width and abundance results, which we discuss in Section~\ref{sec:methods}.

\begin{table}[tbh!]
    \centering
    \begin{tabular}{|l|c|}
    \hline
    \multicolumn{2}{|c|}{\textbf{Identifying Information}} \\ \hline
    \multicolumn{2}{|c|}{{TOI-5882}} \\
    \multicolumn{2}{|c|}{{TIC 232941965}} \\ 
    \multicolumn{2}{|c|}{{Gaia DR3 1869489729418662528}} \\ \hline
    Right Ascension & 311.8886361\,$^\circ$ \\ \hline
    Declination & 34.7374893\,$^\circ$ \\ \hline
    \hline

    \multicolumn{2}{|c|}{\textbf{Spectroscopic Properties (TRES SPC)}} \\ \hline
    {Metallicity ([M/H])} & 0.33 $\pm$ 0.08\,dex \\ \hline
    {Effective Temperature ($T_{\rm eff}$)} & 5723 $\pm$ 85\,K \\ \hline
    {Surface Gravity ($\log{g}$)} & 4.00 $\pm$ 0.12\,dex \\ \hline
     Projected Surface Velocity ($v \sin{(i)}$) & 7.5 $\pm$ 0.4\,$\mathrm{km} \ \mathrm{s}^{-1}$\\ \hline
    {Co-added Signal-to-Noise Ratio} & 58 \\ 
    \hline \hline
    \multicolumn{2}{|c|}{\textbf{Stellar Properties (\exofast{})}} \\ \hline
    {Mass} & $1.334^{+0.055}_{-0.065}$\,$M_{\odot}$ \\ \hline
   {Age} & $4.11^{+0.66}_{-0.52}$\,Gyr\\ \hline
    \hline
    \multicolumn{2}{|c|}{\textbf{Lithium Measurements}} \\ \hline
    {Lithium Equivalent Width} & \ewli{}\\ 
    (Fe 6707.4\,\AA{} uncorrected) & \\ \hline
    {Lithium Equivalent Width} & \ewlife{}\\
    (Fe 6707.4\,\AA{} corrected) & \\ \hline
    {Lithium Abundance A(Li) } & \ali{} \\ \hline
    \end{tabular}
    \caption{TOI-5882 identifying information, spectroscopic properties from the TRES SPC pipeline, stellar properties provided by the highest-probability fit in \cite{2025arXiv250109795V}, and our resulting lithium measurements (see Section~\ref{sec:methods}). }
    \label{tab:5882params}
\end{table}

\subsection{GALAH Control Sample Data}\label{subsec:galah}
Ideally, TOI-5882's lithium abundance would be compared to a coeval control sample drawn from a common moving group or stellar cluster. 
A kinematic analysis within a 50\,pc search radius identified two comoving companions; however, one is a low-mass star with a deep convection zone, and the other a late-stage giant---neither of which provides a meaningful comparison for the lithium abundance of this subgiant star.

Therefore, to place TOI-5882's lithium abundance into proper context, we curated a control sample of field stars with comparable stellar properties (described further in Section~\ref{subsubsec:control}).
Our goal was to construct a comparison set that is homogeneous in data quality and stellar parameters, enabling a meaningful assessment of whether TOI-5882's lithium abundance is anomalous.
We note that this approach makes it challenging to distinguish subtle differences in lithium abundance, as slight variations in stellar age and composition can influence surface lithium abundance. Nonetheless, this approach does enable the identification of strong deviations from typical lithium levels \citep{Soares-Furtado2021}.

While we initially considered using TRES spectra to define our control sample, we found that the number of TRES stars with co-added signal-to-noise ratios ($\mathrm{SNR}>50$) was too limited to support robust statistical conclusions without significantly relaxing our selection criteria. Ideally, we would have constructed the control sample using the same instrument and reduction pipeline as our target star to minimize systematic differences. To preserve the integrity of the comparison, we instead turned to GALactic Archaeology with HERMES (GALAH) DR4 survey \citep{2024arXiv240919858B}, which provides a large spectroscopic dataset with well-characterized stellar parameters, lithium abundances, and equivalent width measurements.
We note that residual inter-pipeline systematics cannot be fully excluded, and we therefore adopt a differential comparison framework and additional photometric constraints to mitigate their impact (see Section~\ref{subsec:li_methods}).
We validate the consistency between GALAH DR4 and TRES lithium abundance measurements for a subigant star observed in both surveys and discuss this cross-check in Section~\ref{subsec:li_methods}. This cross-check ensures that inferences drawn from the GALAH comparison sample are not biased by systematic differences between instruments or pipelines.

The GALAH DR4 survey includes spectra for nearly one million stars, with elemental abundances and stellar parameters (i.e. effective temperature, surface gravity, and metallicity) derived using a reduction pipeline reliant on the spectrum synthesis code \textsc{Spectroscopy Made Easy} (SME; \citealt{Valenti1996,Piskunov2017}) and neural networks trained to map stellar fluxes, stellar parameters, and abundances \citep{2024arXiv240919858B}. 
Lithium equivalent width and abundance are measured from a deblended fit to the Li\,I 6707.8\,\AA\ line using a 3D non-local thermodynamic equilibrium (non-LTE) \textsc{Breidablik} line profile \citep{wang2021,2024MNRAS.528.5394W}. 

\subsubsection{Control Sample Selection}\label{subsubsec:control}
 \begin{figure}[tbh!]
    \centering
    \includegraphics[width=0.95\linewidth]{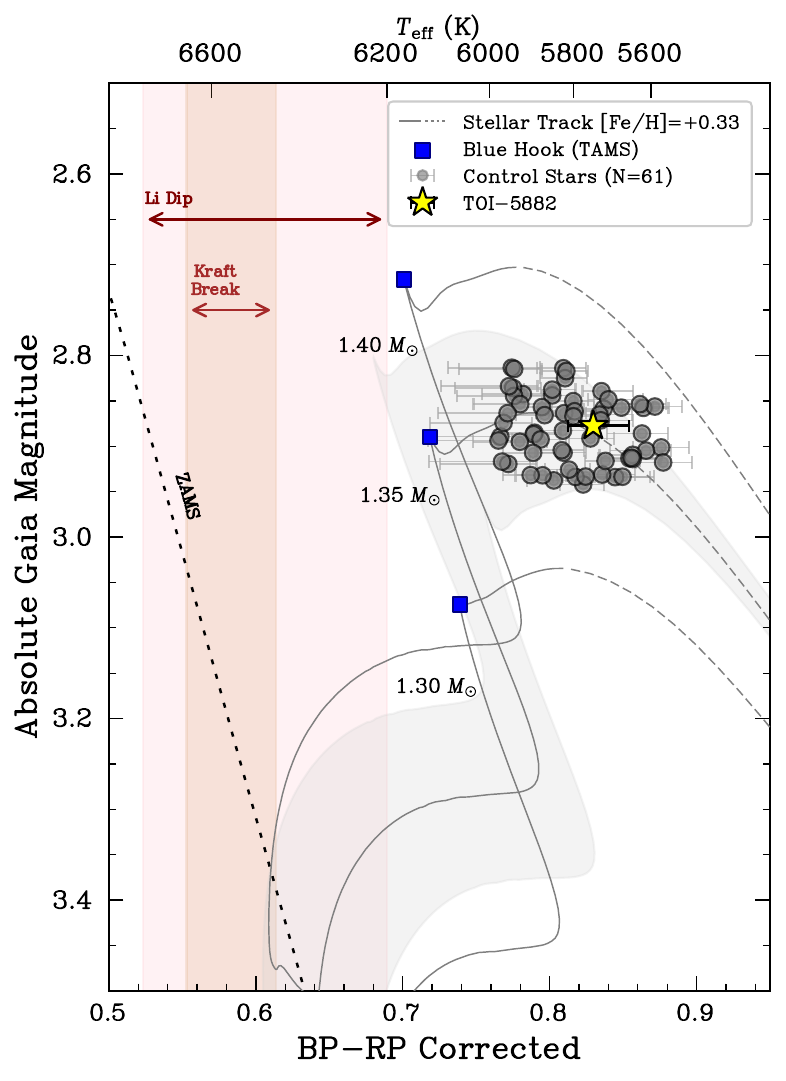}
    \caption{Color-magnitude diagram of TOI-5882 (yellow star) and our GALAH control sample of \populationsize{} stars (grey circles). MIST stellar tracks for $1.35\pm0.05$\,M$_\odot$ are overlaid, with [Fe/H] matching TOI-5882's metallicity. The shaded region represents the [Fe/H] range of $0.33 \pm 0.10$\,dex for $1.35$\,M$_\odot$. The dotted line illustrates the location of the zero-age-main-sequence for a 0.33 dex population. 
    Dashed segments indicate post-main-sequence evolution. TOI-5882 lies on the subgiant branch, consistent with early post-main-sequence evolution. The location of the blue hook (blue squares), Kraft break \citep{2024ApJ...973...28B}, and lithium dip are shown for added context.}
    \label{fig:galah_control}
\end{figure}

To curate our sample, we drew targets from the  \texttt{galah\_dr4\_allstar\_240705.fits} catalog, which is recommended for science cases utilizing stellar parameters and abundances.\footnote{\url{https://www.galah-survey.org/dr4/the_catalogues}} 
For added data quality assurance, we followed the recommendations outlined on the GALAH Best Practices webpage.\footnote{\url{https://www.galah-survey.org/dr4/overview/}; \\ \url{https://www.galah-survey.org/dr4/using_the_data/}}
More specifically, we only included sources with the recommended flag settings for the following parameters: \texttt{snr\_px\_ccd3 > 30} (red camera signal-to-noise per pixel), \texttt{flag\_red==0} (indicating successful data reduction), \texttt{flag\_sp==0} (removes sources with missing data, very low signal spectra, poor analysis results, and signs of binarity), and \texttt{flag\_fe\_h==0} (indicating a reliable metallicity).
Since our investigation included a lithium equivalent width analysis, we set lithium elemental abundance flag to \texttt{flag\_a\_li$\leq$1}, which prevented the culling of stars with low lithium equivalent width measurements. 
The measured A(Li) values for these lithium-poor targets are reported as upper limits.

Further, we wanted to ensure accurate calibration of each control star's position on a color-magnitude diagram (absolute magnitude versus reddening-corrected color, accounting for distance varying line-of-sight dust conditions) using measurements from \textit{Gaia} Data Release 3 (DR3, \citealt{GaiaDR3}).
Therefore, we included only stars with measured parallax, ($A_{\textrm{G}}$) extinction, and reddening ($E(BP-RP)$) measurements.
To ensure well-constrained distances, we restrict our sample to sources with \texttt{parallax\_over\_error} ratios greater than five.
Lastly, to minimize further contamination from unresolved binaries (beyond what was mitigated by the \texttt{flag\_sp==0} setting), we include only targets with Renormalized Unit Weight Error (RUWE) values below 1.25. This threshold, a reduced $\chi^2$ statistic, has been shown to effectively exclude unresolved stars in multiple systems \citep{Belokurov2020}. 

To construct a reliable and robust control sample, we selected stars with GALAH-reported stellar properties that closely match those of TOI-5882 (as provided in Table~\ref{tab:5882params}).
We selected stars using the following parameters ranges:

\begin{itemize}
    \item $\mathrm{[Fe/H]}$ within $\pm 0.1$\,dex of our target TOI-5882. 
    Control star metallicities were drawn from GALAH DR4, which reports $\mathrm{[Fe/H]}$, while TOI-5882's metallicity was measured as $\mathrm{[M/H]}$ using the TRES SPC pipeline. Although $\mathrm{[M/H]}$ and $\mathrm{[Fe/H]}$ can diverge in stars with non-solar abundance patterns, TOI-5882 is a metal-rich star with a well-constrained age of 4\,Gyr (see Table~\ref{tab:5882params}) and kinematics consistent with thin disk membership. For such stars, the difference between $\mathrm{[M/H]}$ and $\mathrm{[Fe/H]}$ is typically negligible \citep{Mikolaitis2017}, making this approximation appropriate for our control sample selection.
    \item Surface gravity, $\log{g}$, within $\pm 0.2$\,dex of the target. Values for control stars were obtained from GALAH DR4, while TOI-5882's $\log{g}$ was measured using the TRES SPC pipeline. 
    \item $T_\mathrm{eff}$ within $\pm 250$\,K of the target. While TOI-5882's effective temperature was measured as $5723 \pm 85$\,K using the TRES SPC pipeline, an independent SED fit yielded a higher value of $5920 \pm 210$\,K \citep{2025arXiv250109795V}. Our control stars fall within the combined uncertainty bounds of the target's temperature estimates. Control star temperatures were drawn from GALAH DR4.  
    \item \textit{Gaia} reddening-corrected color within $\pm 0.065$\,mag of the target (using the $G_{\mathrm{BP}}-G_{\mathrm{RP}}$ bandpass). This offers a nearly model-independent proxy for temperature that is less sensitive to spectral fitting assumptions.
    \item Absolute \textit{Gaia} magnitude within $\pm 0.065$\,mag of the target  (parallax and extinction corrected). 
    \end{itemize}

To ensure our control sample represents evolved subgiants rather than young, lithium-rich stars, we assessed each star for potential youth contamination. 
We searched for comoving companions of each star within 25\,pc using \textit{Gaia} DR3 astrometry and applied kinematic cuts via the \textsc{Friendfinder} tool \citep{Tofflemire2021}, setting tangential and radial velocity offset constraints of $<5$\,km\,s$^{-1}$ of predictions. 
For stars with identified comoving ensembles of five or more stars, we measured the age of the system using the photometric variability framework from \citet{Barber2023}.
This analysis identified one young star in the control sample (Gaia\,DR3\,3044345872508615680), indicating primordial lithium retention rather than post-main-sequence enrichment. 
We measured an age of $\tau = 405^{+196}_{-106}$\,Myr for the ensemble of 44 comoving members. 
We removed this star from our control sample.

These criteria yielded a broad enough sample to support a robust control group in the post-main-sequence phase. This resulted in a sample size of \populationsize{} control stars: 27 stars with \texttt{flag\_a\_li==0} (definitive A(Li) measurements) and 34 stars with \texttt{flag\_a\_li==1} (upper limits on A(Li)).
In Figure~\ref{fig:galah_control}, we illustrate the color-magnitude diagram (CMD) position of TOI-5882 (yellow star) and our resulting control sample (black points). The mapping between \textit{Gaia} bandpasses and effective temperature comes is adopted from \cite{Pecaut2013}. 
Our target star is shown to be well-embedded in the control sample population.

All of the plotted stars have been corrected for parallax, extinction, and reddening.
Also plotted are evolutionary tracks from the Mesa Isochrones and Stellar Tracks (MIST) repository \citep{Choi_2016}, computed using the Modules for Experiments in Stellar Astrophysics (MESA; \citealt{Dotter_2016,Paxton:2011, Paxton:2013, Paxton:2015}) code. 
To allow for general comparisons between our representative stellar population, our MESA models do not incorporate more physically complex overshoot mixing or wind loss prescriptions. 
We note that the MIST models shown here and those used in our analysis include convective overshoot (see Appendix~\ref{appendix_inlist} for a complete list of our inlist parameters), however, this represents a specific choice of overshoot prescription. 
Alternative implementations can shift and broaden the terminal-age main-sequence blue hook in color--magnitude space (see Figure~\ref{fig:galah_control}). 
However, our control sample is defined entirely using directly observed quantities, rather than inferred stellar mass or evolutionary phase. As a result, any uncertainty in the location of the blue hook and lithium dip does not affect the sample construction or the differential lithium comparison.
Moreover, our matching protocol ensures that control stars share a similar main-sequence thermal history with TOI-5882, including any Li Dip depletion, so the observed differential enhancement cannot be attributed to Li Dip effects not equally present in the reference population.

The evolutionary tracks correspond to a narrow mass range of $1.35\pm0.05$\,\Msun{} and were chosen to reflect the metallicity of our target star (as listed in Table~\ref{tab:5882params}). 
The stellar tracks are depicted with dashed lines when the star is no longer undergoing hydrogen core-fusion (post-main-sequence stage).
A shaded region around the $1.35$\,\Msun{} track represents the [Fe/H] range of $0.33 \pm 0.10$\,dex.
TOI-5882 is shown to overlap with the subgiant branch, one of the earliest stages of post-main-sequence evolution.

We also note that the projected rotational velocity of TOI-5882 (\vsini$ = 7.5 \pm\,0.4$ km\,s$^{-1}$) is typical of the control sample (median \vsini$=6.1$\,km\,s$^{-1}$, mean \vsini$=6.8$ km s$^{-1}$), indicating that it is not an extreme rotator relative to the comparison stars.

\section{Lithium Signature Measurement and Analysis}
\label{sec:methods}
\subsection{Lithium Equivalent Width and Abundance Measurement}
\label{subsec:li_methods}

\begin{figure}[tbh!]
    \centering
    \includegraphics[width=0.97\linewidth]{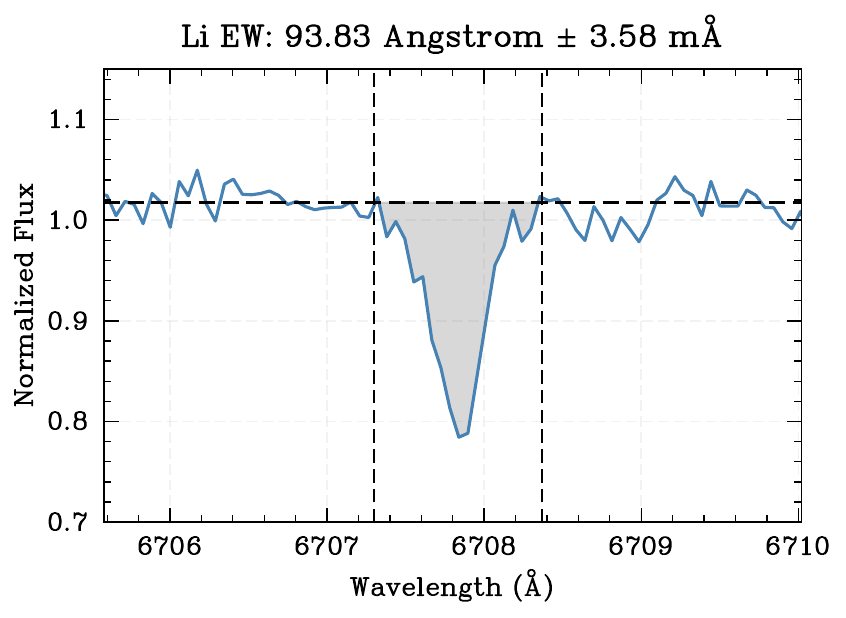}
    \caption{Li\,I feature of TOI-5882 centered at 6707.8\,\AA{}, as seen in the co-added TRES spectra. Continuum is set at 1.0174 with measurement boundaries at 6707.30\,\AA{} and 6708.37\,\AA{}. We determined an Li\,I EW measurement of \ewli{} (not accounting for the Fe feature at 6707.4\,\AA{}). The Fe-corrected value is provided in Table~\ref{tab:5882params}}
    \label{fig:Li_EW}
\end{figure}

We measured the Li\,I EW of the resonant doublet Li\,I feature centered at 6707.8\,\AA{} using our co-added TRES spectra of TOI-5882 as seen in Figure~\ref{fig:Li_EW}.
The Li\,I absorption feature was easily distinguishable with the high resolution of the TRES spectrograph and our high co-added SNR of 58.

For this procedure, we employed the software tool \texttt{Specutils} \citep{specutils,nicholas_earl_2024_10681408}. We fit the continuum with \texttt{fitting.fit\_generic\_continuum} and measured the Li\,I EW with \texttt{analysis.equivalent\_width}. As shown in Figure~\ref{fig:Li_EW}, we set the continuum at 1.0174 and imposed measurement boundaries at 6707.30\,\AA{} and 6708.37\,\AA{}.
This resulted in an Li\,I EW of \ewli{} for TOI-5882. The uncertainty is calculated following \citet{ebbets_1995} Equation~5, which we rewrite as:
\vspace{0.6cm}
\begin{equation}
\label{eq:ew_unc}
\sigma_{\mathrm{Li\,I EW}}^2 = \left( \frac{N \cdot \left( \frac{\Delta\lambda}{N} \right)^2}{\mathrm{SNR}^2} \right) \cdot \left( \frac{\min(\mathrm{flux})}{\max(\mathrm{flux})} \right),
\end{equation}
where N is the number of points within the measurement range, $\Delta\lambda$ is the difference between the Li\,I EW wavelength measurement edges, SNR is the co-added signal-to-noise ratio for our wavelength order, and flux is the flux values within the Li\,I EW measurement bounds. 
This initial measurement does not account for the Fe feature at 6707.40\,\AA{}; we describe the process by which we subtract the iron blend contribution later in this section.

We then converted our Li\,I EW measurement to an iron-corrected equivalent width and logarithmic lithium abundance, A(Li), as measured on the 12-point scale.\footnote{The A(Li) measurement is a standardized way of reporting lithium abundances in stellar atmospheres on a logarithmic scale relative to hydrogen: $\mathrm{A(Li)}=   \log \left( \frac{N_{\mathrm{Li}}}{N_{\mathrm{H}}} \right) + 12\,\mathrm{dex},$
where $N_{\mathrm{Li}}$ is the number of lithium atoms and $N_{\mathrm{H}}$ is the number of hydrogen atoms.
The scale accounts for the fact that the least abundant chemical species are found in ratios of $1:10^{12}$ H atoms.} Our A(Li) conversion was performed using curves of growth from \cite{Franciosini_2022}.
This step incorporates the iron abundance, effective temperature, and surface gravity measurements, as determined by the TRES SPC analysis (described in Section~\ref{subsec:tres}). 
Thus, A(Li) abundances are more model-dependent observables than equivalent widths, as they incorporate model atmospheres, line lists, and NLTE corrections. This consideration informed our decision to also perform an equivalent widths comparison (described in Section \ref{subsec:li_results}).

We accounted for the iron blend by applying the A(Fe) corrections provided in \cite{Franciosini_2022}, using the TRES SPC [M/H] measurement of TOI-5882, assuming an iron abundance is the same as the SPC metallicity. The curves of growth of the Fe\,I 6707.43\,\AA{} are computed from a grid of Model Atmospheres with a Radiative and Convective Scheme (MARCS) synthetic spectra (\citealt{Gustafsson_2008}), following the method of \cite{de_Laverny_2013}. 
We estimate the Li\,I EW of the Fe\,I blended line by interpolating across Table~A.2 of \cite{Franciosini_2022} in order to match the observed properties of TOI-5882. 
We then subtracted this estimated Fe\,I EW from our directly measured Li\,I EW (as listed in Table~\ref{tab:5882params}) in order to determine the equivalent width of the Li\,I~6707.8\,\AA{} feature alone. With this correction, we find a final, deblended Li\,I EW of \ewlife{} for TOI-5882. 
We computed our A(Li) using this deblended Li\,I EW and curves of growth from \cite{Franciosini_2022}.
As provided in Table~\ref{tab:5882params}, our lithium abundance was measured as A(Li)=\ali{}.  

While the uncertainty in our Li\,I EW measurement obtained using Equation~\ref{eq:ew_unc} yields an error in A(Li) of $\pm 0.028$\,dex, we also must account for uncertainties in the stellar input parameters from the TRES SPC pipeline.
To optimally account for this error, we varied these stellar parameters across their respective extremes, measuring the corresponding A(Li) and associated error.
This approach provided a more accurate representation of the true underlying uncertainty, which is larger than the $0.028$\,dex uncertainty estimate indicated by our initial Li\,I EW analysis. 
As a result, we adopt a more conservative A(Li) error limit of $\pm 0.12$\,dex for the remainder of this work.

There is conceivable concern that the comparison of lithium equivalent width measurements between our TRES spectral analysis and GALAH DR4 analyzed control stars could be inherently inconsistent given the different instruments and reduction pipelines.
To confirm consistency in our investigation, we analyzed the Li\,I EW measurement of TOI-844 (TIC\,380886535,
Gaia\,DR3\,2534988933319166976)---one of the rare subgiant stars observed by both the TRES and GALAH instruments. 
While limited to a single overlapping object, TOI-844 closely matches the median effective temperature and $v\sin i$ of the control sample and occupies the same evolutionary phase, making it a representative cross-check.
A larger sample of subgiant comparisons could not be obtained due to lack of existing GALAH DR4 measurements, SNR limitations in the TRES survey, a lack of TRES targets with distinguishable lithium features, and declination limit differences across the two instruments in differing hemispheres. 

We measured the lithium feature at 6707.8\,\AA{} for the test star, as we had done for TOI-5882.
This enabled us to appropriately define the continuum and place the equivalent width bounds. 
Accounting for the iron feature at $6707.43$\,\AA{}, we find our corresponding Li\,I EW to be within 5\,m\AA{} (which is within our measurement uncertainty) of the measurement provided by GALAH DR4. 
The Li\,I EW agreement supports the conclusion that the lithium equivalent width derived for TOI-5882 from our TRES observations is consistent with those measured in the GALAH control sample.
It is also worth noting that the $T_{\mathrm{eff}}$ and $\log{g}$ values calculated using GALAH spectra were within the measurement and error bounds of the values measured using the TRES SPC process, supporting the consistency between these two data products. 

\subsection{Comparison with the GALAH Control Sample}
\label{subsec:li_results}

Comparing the Li\,I EW measurement of TOI-5882 with our GALAH DR4 control sample, we find that our target star is in the \percentile{} percentile, indicating that it is more enhanced than all but a single control star. 
To more rigorously quantify this statistical rarity, we performed a tail probability analysis: among the \populationsize{} control stars, 60 have EW(Li)\,$<$\,\ewlife{}, yielding an empirical tail probability of $P(X \geq$ \ewlife{}$) = 0.016$ (1.6\%). 
We then used a non-parametric bootstrap analysis with replacement
to assess the robustness of our result, accounting for measurement uncertainties in both the target and control samples. 
For each of 100,000 iterations, we resampled the control dataset, adding Gaussian noise consistent with the median measurement errors, and drew a random realization of the target. 
The resulting distribution yielded a median percentile rank of 99.09\% with a 95\% confidence interval of $^{+0.91}_{-3.64}$, supporting our prior conclusion that TOI-5882 is lithium enriched in this regime.

\begin{figure}[tbh!]
  \centering
  \begin{subfigure}[tbh!]{0.47\textwidth}
    \centering
    \includegraphics[width=\textwidth]{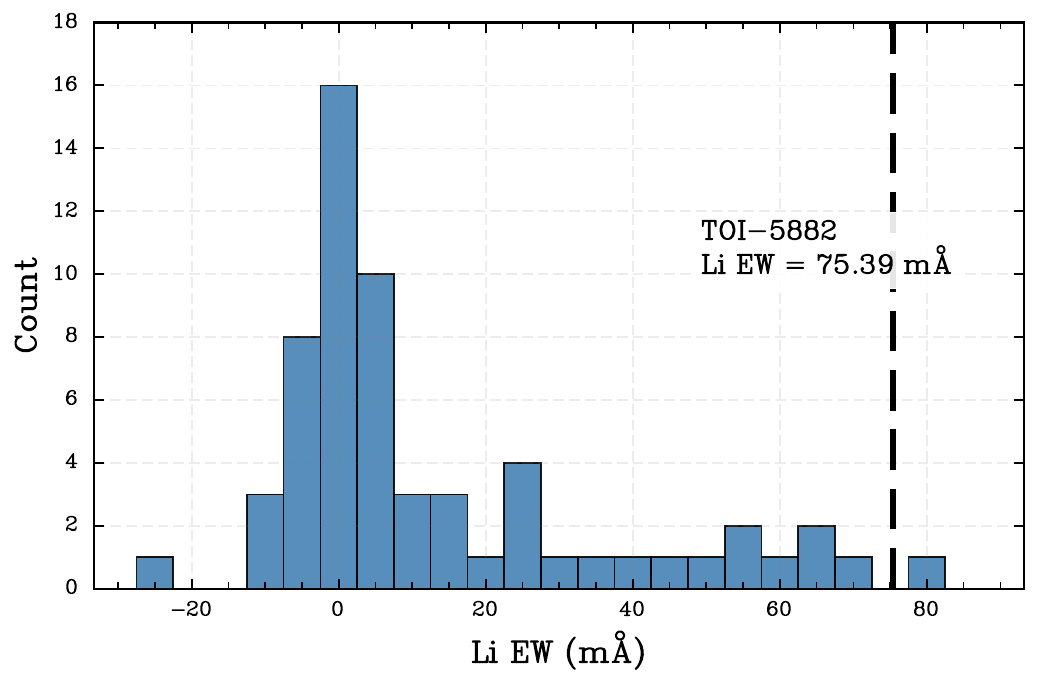}
    \label{fig:ew_hist}
  \end{subfigure}
  \hfill
  \begin{subfigure}[tbh!]{0.47\textwidth}
    \centering
    \includegraphics[width=\textwidth]{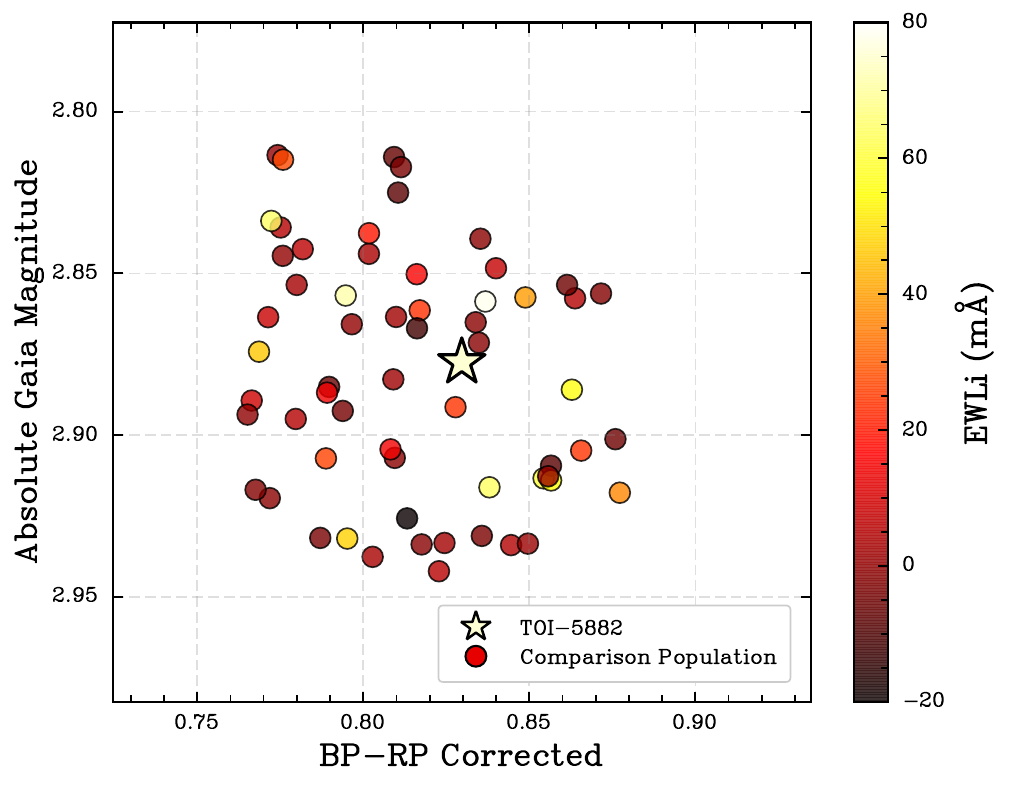}
    \label{fig:ew_scatter}
  \end{subfigure}
    \caption{\textbf{Top Panel:} Li\,I EW histogram of \populationsize{} GALAH control stars. TOI-5882 (dashed line at \ewlife{}) is exceeded by only one control star, ranking in the \percentile{} percentile. \textbf{Bottom Panel:} Control sample Li\,I EW values (color-coded circles) compared with TOI-5882 (star) in CMD space, showing TOI-5882 occupies typical subgiant CMD position while exhibiting extreme lithium enhancement.}
  \label{fig:ali_ew_twopanel}
\end{figure}

The histogram of Li\,I EW width measurements pertaining to our target and control sample is shown in the top panel of Figure~\ref{fig:ali_ew_twopanel}. 
The control sample Li\,I EW strengths range between -20 to 80\,m\AA{}, where negative values correspond to noise in the continuum among extremely lithium-poor targets. Most of the comparison stars show low Li\,I EW abundances. 
Near the far-right edge of the distribution is a black dashed line, indicating the strength of the Li\,I EW measurement of TOI-5882.
TOI-5882 belongs to the extreme tail of this skewed distribution, not the broader moderately-enhanced population. 
The one star with a larger lithium equivalent width (78.2\,m\AA{}) demonstrates that even more extreme cases of lithium enrichment exist and similarly warrant detailed investigation.

TOI-5882 also ranks in the \percentile{} percentile when comparing derived A(Li) values (accounting for upper limits among lithium-poor stars). 
Though in this analysis, we emphasize Li\,I EW measurements because they are directly observable and independent of model atmospheres, which can appreciably alter the measurement (generally by several tenths of a dex; \citealt{Sun_2025}). 
As shown in the bottom panel of Figure~\ref{fig:ali_ew_twopanel}, TOI-5882 occupies the same CMD space as the control population, yet exhibits a lithium level exceeded by only one star in the sample.
Additionally, we examined the relationship between Li\,I EW and metallicity across the control sample, motivated by the expectation that metallicity can influence stellar structure and line formation, and could introduce systematic trends in measured lithium equivalent widths. We find no statistically significant correlation.

While our cross-validation demonstrates consistency between TRES and GALAH measurements (Section~\ref{subsec:li_methods}), future work would benefit from a homogeneous spectroscopic survey of subgiants using a single instrument and reduction pipeline to eliminate residual systematic uncertainties in stellar parameters.
As we will show in Section~\ref{subsec:li_methods}, the residual systematics discussed are individually expected to operate at the level of a few to tens of m\AA, while TOI-5882's enhancement of $\sim$57~m\AA\ above the control sample median is large enough that no single identified systematic, nor any plausible combination, is expected to bridge this gap.
Additionally, curating a larger comparison population with even tighter constraints on stellar parameters and CMD position would allow for a more precise assessment of lithium enrichment, strengthening any inference of anomalous surface abundances.
We also note that atomic diffusion processes, including gravitational settling and radiative levitation, are active in this temperature regime; while the tight parameter-space matching mitigates systematic offsets, differential diffusion effects within the sample cannot be fully excluded and represent a residual uncertainty in the comparison.

\subsection{Robustness to Control Sample Definition}
\label{subsec:robustness}

A key aspect of this analysis is the definition of a comparison sample of stars with similar fundamental properties. 
To assess the sensitivity of our results to this choice, we systematically vary the selection criteria used to define control stars and evaluate the resulting lithium distributions.

\begin{figure*}
    \centering
    \includegraphics[width=\linewidth]{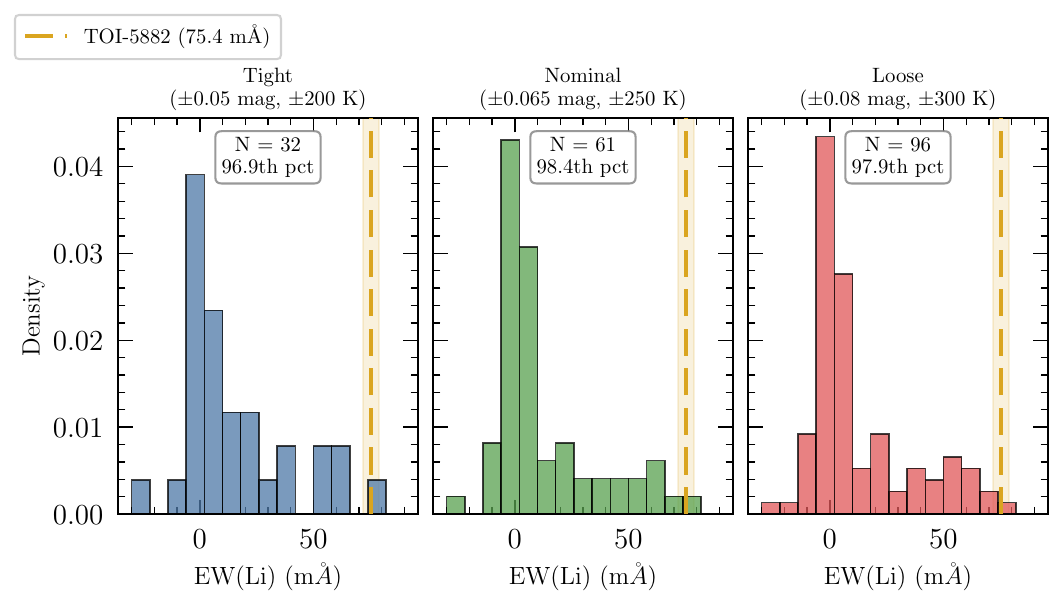}
    \caption{Lithium equivalent width distributions  for three comparison samples constructed using tight, nominal, and loose constraints in color--magnitude space and $T_{\rm eff}$. 
    In all cases, the distributions are strongly peaked at low Li\,I EW , while TOI-5882 lies in the extreme upper tail. 
    The inferred percentile rank remains consistently high despite the change in sample size, demonstrating that the identification of TOI-5882 as lithium-enhanced is not acutely sensitive to the specific choice of constraints.}
    \label{fig:histogram}
\end{figure*}

We first construct three representative samples using tight, nominal, and loose windows in color--magnitude space and $T_{\rm eff}$. 
Figure~\ref{fig:histogram} illustrates the resulting Li\,I EW distributions for each case. 
Across all selections, the distributions are strongly peaked at low equivalent widths, while TOI-5882 consistently lies in the extreme upper tail. 
Despite the increase in sample size from $N=32$ to $N=96$, the inferred percentile rank of TOI-5882 remains high (96.9th, 98.4th, and 97.9th percentile, respectively), indicating that the identification of TOI-5882 as a lithium-enhanced outlier is not driven by a particular choice of selection window.

\begin{figure*}
    \centering
    \includegraphics[width=\linewidth]{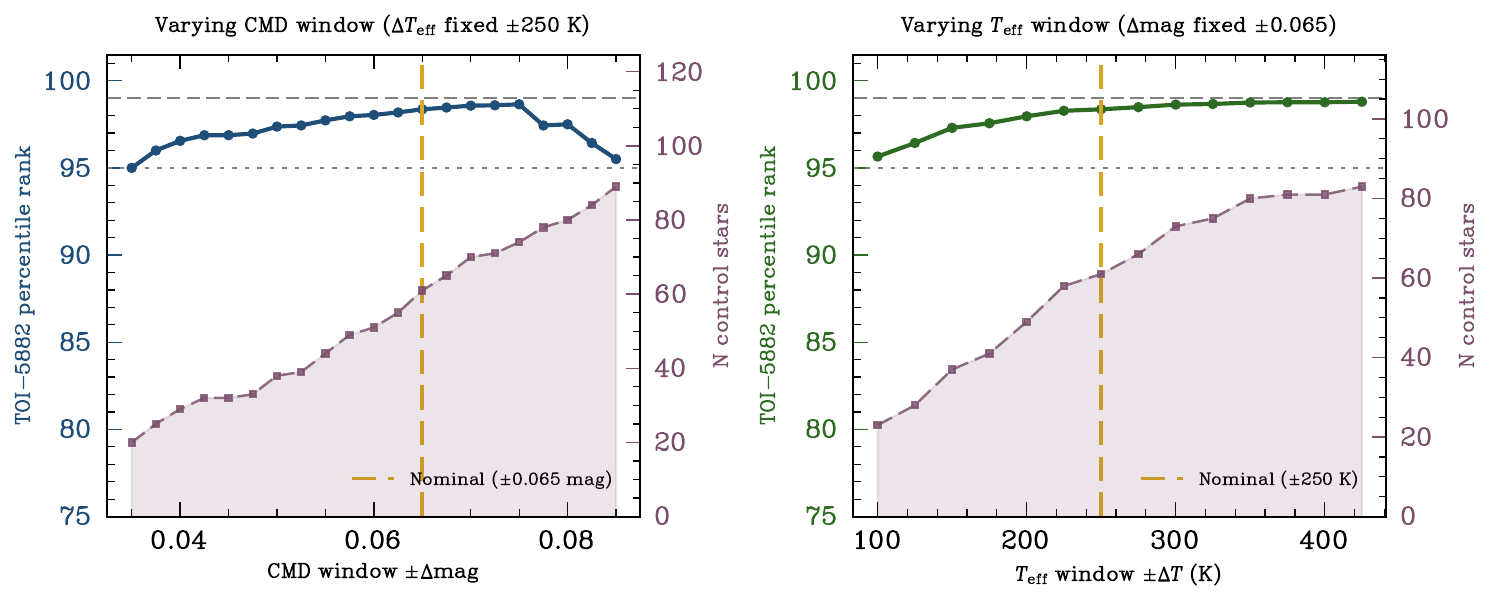}
    \caption{Percentile rank of TOI-5882 as a function of the adopted selection window. 
    Left: variation with color--magnitude (CMD) window width. 
    Right: variation with $T_{\rm eff}$ window width. 
    The percentile rank remains above $\sim$95\% across all tested configurations, peaking above $\sim$98\% for intermediate window sizes. 
    The modest decline at large CMD widths reflects the inclusion of less well-matched comparison stars, while the $T_{\rm eff}$ selection shows minimal impact beyond $\pm$200\,K.}
    \label{fig:percentile}
\end{figure*}

In addition, we vary the CMD and $T_{\rm eff}$ windows independently and track the resulting percentile rank (Figure~\ref{fig:percentile}).
For the CMD sweep, the percentile rank rises from $\sim$95\% at the narrowest windows, where the sample size is small, peaks above $\sim$98\% for window widths of $\pm$0.055--0.075\,mag, and declines to $\sim$96\% at the widest windows as less well-matched stars are included. 
For the $T_{\rm eff}$ sweep, the percentile rank stabilizes above $\sim$98\% once the window reaches $\pm$200\,K and remains effectively flat at larger widths. 
This behavior indicates that the $T_{\rm eff}$ selection is not a dominant source of uncertainty.

Finally, we include a direct spectrum-to-spectrum comparison in Appendix~\ref{appendix_spectra}, illustrating the Li\,I EW region for TOI-5882 alongside four control stars. 
The enhanced depth of the Li\,I EW feature in TOI-5882 is directly evident in the normalized spectra, and the strongest control star provides a close match to the feature depth, with the Li\,I EW strength consistent within measurement uncertainties. 
This comparison confirms that the observed enhancement is not an artifact of equivalent width measurement or survey-specific systematics.
Taken together, these tests demonstrate that the lithium enhancement in TOI-5882 is robust to reasonable variations in control sample definition and represents a genuine outlier relative to the underlying population.

\section{Implications for Planetary Engulfment}
\label{sec:discussion}

TOI-5882 exhibits extreme lithium enhancement (\percentile{} percentile) and hosts a massive brown dwarf companion ($22.01^{+0.61}_{-0.72}$\,\MJ{}, $P = 7.1$\,d) in precisely the evolutionary regime where \citet{Soares-Furtado2021} predicted engulfment signatures would be detectable. Before quantitatively evaluating the engulfment hypothesis, we consider alternative explanations.

Internal lithium production is effectively ruled out by evolutionary state. The Cameron-Fowler mechanism has been shown to operate effectively after the luminosity bump on the red giant branch. TOI-5882's CMD position (Figure~\ref{fig:galah_control}) confirms it is an early subgiant (mass $1.334^{+0.055}_{-0.065}$\,$M_\odot$, age $4.11^{+0.66}_{-0.52}$\,Gyr; \citealt{2025arXiv250109795V}) where this mechanism is not expected to operate.

Primordial lithium retention is excluded by multiple observational diagnostics. This is critical because young pre-main-sequence stars can occupy CMD positions degenerate with evolved subgiants, making photometric selection insufficient. 
This was made evident in our youth contamination assessment of the control sample (Section~\ref{subsubsec:control}), which identified one such young contaminant.
Ground-based spectroscopic and photometric follow-up reveals that TOI-5882 displays no youth indicators. This includes no confident signs of infrared excess or H$\alpha$ emission, no affiliation with a young comoving ensemble, and a rotation rate that is commensurate with its maturity \citep{2025arXiv250109795V}.

With alternative enrichment mechanisms eliminated, we evaluate whether planetary engulfment can quantitatively account for TOI-5882's lithium excess relative to the control sample baseline.
As discussed in Section~\ref{sec:intro}, the brown dwarf companion provides a dynamical mechanism for perturbing inner planets onto star-grazing orbits through secular perturbations and resonances \citep[e.g.,][]{2005ApJ...627.1001T, 2008ApJ...686..621F, 2011ApJ...732...74G, Steffen2012, Mustill2015}.

The critical test is whether a plausible planetary mass can deliver sufficient lithium to produce the observed enhancement. We evaluate this in three steps: (1) determining \textit{where} in the stellar interior planetary material would be deposited (Section~\ref{subsec:disassociation}), (2) calculating the required mass assuming proto-solar composition following standard methodology \citep{Montalban2002, Soares-Furtado2021} (Section~\ref{subsubsec:protosolar}), and (3) recalculating the required mass using more realistic planetary compositions that account for the heavy-element enrichment of planetary bodies (Section~\ref{subsubsec:realistic}). Throughout this analysis, we use ``planetary engulfment" to refer to the hypothesized event, notwithstanding the formal planet/brown dwarf mass boundary.

\subsection{Planetary Engulfment \& Convective Zone Deposition}
\label{subsec:disassociation}

The observability of engulfment-derived chemical signatures depends critically on \textit{where} planetary material is deposited. 
Stars of TOI-5882's mass and evolutionary stage are radiative in their interior and convective near their surface. When planetary material is deposited anywhere in the convective zone, convection rapidly mixes it throughout, altering the observable surface abundance. Material deposited below the convective zone base remains hidden from spectroscopic detection.

Using the Modules for Experiments in Stellar Astrophysics (\textsc{MESA} version 24.08.1; \citealt{Dotter_2016, Paxton:2011, Paxton:2013, Paxton:2015}), we modeled TOI-5882 as a non-rotating $1.35$\,$M_{\odot}$ star with $\left[\text{Fe}/\text{H}\right] = 0.33$\,dex (Table~\ref{tab:5882params}). Model parameters are provided in Appendix~\ref{appendix_inlist}. 
We estimate a convective zone mass of $0.035 \pm 0.033$\,\Msun{} (averaging to 37 $M_{\rm J}$) and a convective zone depth of $0.42 \pm 0.23$\,\Rsun{} for this target.
Uncertainties reflect the range of viable mixing-length parameters ($\alpha_{\rm MLT} = 1.65$--$1.80$) for stars with similar $\log g$, $T_{\rm eff}$, and [Fe/H] \citep{viani2018}, plus the $\pm 1\sigma$ uncertainty in TOI-5882's effective temperature (Table~\ref{tab:5882params}).

While planetary material deposited anywhere in the convective zone rapidly mixes and changes the surface abundance, mixing processes such as convective overshooting and thermohaline mixing gradually transfer planetary material from the convective zone into the radiative zone \citep[e.g.,][]{Vauclair2004,Garaud2011}. Over time, this will bring the surface abundances closer to their pre-engulfment value. The efficiency of these mixing processes is uncertain, and therefore so is the timescale over which the surface engulfment signatures fade.

Planets with the lowest structural integrity will be tidally disrupted above the stellar surface. Depending on their mass, the accretion rate onto the star may be super-Eddington, resulting in an outflow in which a significant fraction of the planetary material becomes unbound instead of being accreted \citep{Metzger2012}. In contrast, high-density planets may remain intact as they inspiral below the base of the convective zone, likely depositing some but not all of their material in the convective zone.

The location within the star where the planet is destroyed depends on the properties of the star and the planet. Given that the properties of the star change over timescales comparable to those over which chemical signatures vanish ($\lesssim1$\,Gyr; \citealt{Soares-Furtado2021,sevilla_long_term_lithium_signatures,aida_timelines}), we consider both the present-day properties of TOI-5882 and those 1\,Gyr in the past. The planet will be tidally disrupted before reaching the surface of the star if $\rho_p\lesssim5\rho_\star$, where $\rho_p$ is the planetary density and $\rho_\star$ is the stellar density \citep{Metzger2012}. 
The present-day density of TOI-5882, estimated to be $\rho_\star=0.8 \pm 0.2$\,g/cc with \textsc{MESA} models, 
indicates that even planets with compact, volatile-poor planets (those with rocky/metal-rich compositions and modest H/He envelopes) are expected to be destroyed by tidal forces before reaching the stellar surface. According to statistical planet mass-radius relations \citep{Chen2017}, the masses of such planets typically lie in the range \( 7\,M_\oplus\lesssim M_\mathrm{p}\lesssim 1\,M_\text{J} \). 

However, these mass-radius relations might not accurately represent the properties of the planet at the onset of the merger.
For example, if tidal dissipation is responsible for the merger, the increased flux on the planet over the course of many orbital periods might inflate the planet and make it more vulnerable to tidal disruption \citep[e.g.,][]{Bodenheimer2001,Millholland2019}. 
In contrast, a dynamical event that launches the planet directly into the star could do so without significantly altering the structure of the planet.\footnote{When we discuss the properties of the planet, we mean the properties of the planet at the time of the merger.} 

If we extend the tidal disruption criterion to the interior of the star, we can approximate the destruction location of ingested companions. While an engulfed brown dwarf is likely to be disrupted below the base of the convective zone \citep{Yarza2023}, it is expected that ingested companions will gradually lose mass as a result of ram pressure ablation. As a result, some of the companion's mass will be mixed into convective zone even if the companion is ultimately destroyed below this region. In Section~\ref{subsec:engulfed_mass}, we estimate the total mass of the planetary material that may have ended up in the convective zone of TOI-5882, which may not be the same as the entire mass of the planet.
Given these uncertainties, we tentatively favor the accretion of a planet---as opposed to a brown dwarf---in order to deposit material within the convective zone.

Engulfment at an earlier stage of evolution facilitates the deposition of material into the convective zone because the star is, on average, denser, resulting in the disruption of planetary companions at larger separations. However, when accounting for engulfment that took place 1\,Gyr ago, our conclusions remain similar as in the present-day case. 
Our order-of-magnitude estimates suggest brown dwarfs will survive below the base of the convective zone, indicating that the enhanced lithium abundance of TOI-5882 is likely derived from an engulfed planetary companion rather than the partial dissolution of a brown dwarf.
The engulfment mass estimates we present in Section~\ref{subsec:engulfed_mass} offer further support of this conclusion. 

\subsection{Planetary Mass Estimates}\label{subsec:engulfed_mass}
\subsubsection{Proto-Solar Composition Assumption}
\label{subsubsec:protosolar}
To evaluate whether the lithium enrichment observed in TOI-5882 can be explained by the engulfment of a planetary-mass companion, we compute the mass of a planet required to reproduce the present-day stellar lithium abundance within the convective zone.
As described in Section~\ref{subsec:disassociation}, using our aforementioned \textsc{MESA} models (which vary viable $\alpha$ mixing length parameters and the $\pm1\sigma$ bounds of TOI-5882's effective temperature), we estimate a convective zone mass of $\left(6.96\pm  6.65\right)\times10^{31}$\,grams, which averages to 37\,\MJ{} where \MJ{}$= 1.9 \times 10^{30}$\,g is the mass of Jupiter.
\textsc{MESA} also provides an estimated hydrogen mass fraction in the convective zone of $X_\text{CZ}=0.65$, so that the estimated total number of hydrogen atoms in the CZ is:
\begin{equation}
    N_\text{H,CZ} = \frac{X_\text{CZ} M_\text{CZ}}{m_\text{H}} = \left( 2.70 \pm 2.58 \right) \times 10^{55}\,\text{atoms},
\end{equation}
where the mass of a hydrogen atom is $m_\text{H} = 1.7 \times 10^{-24}$\,g.
Following the methodology of \citet{Soares-Furtado2021}, we assert that the amount of lithium and hydrogen in the convective zone comes from two components:
\begin{align}
    N_\text{CZ,Li} &= N_\text{s,Li} + N_\text{p,Li}\\
    N_\text{CZ,H} &= N_\text{s,H} + N_\text{p,H} 
\end{align}
where $N_\text{s,Li}$ and $N_\text{s,H}$ are the pre-engulfment stellar components and $N_\text{p,Li}$ and $N_\text{p,H}$ are the engulfed planetary components of lithium and hydrogen, respectively.
Thus, the enhanced lithium signature within the convective zone from a planetary engulfment event can then be estimated as 
\begin{equation}
    10^{A(\text{Li})_\text{CZ}-12} = \frac{N_\text{CZ,Li}}{N_\text{CZ,H}} = \frac{N_\text{s,Li} + N_\text{p,Li}}{N_\text{s,H} + N_\text{p,H}},
\end{equation}
where $A(\text{Li})_\text{CZ} = 2.5 \pm 0.1$\,dex is the relative lithium abundance in the convective zone of TOI-5882 as inferred from our spectral analysis (see Section~\ref{subsec:li_methods}).
Similarly, the relative lithium abundance of the star pre-engulfment is 
\begin{equation}
    10^{A(\text{Li})_\text{s}-12} = \frac{N_\text{s,Li}}{N_\text{s,H}},
\end{equation}
where $A\text{(Li)}_{\text{s}} = 1.42^{+0.06}_{-0.02}$\,dex is the baseline stellar lithium abundance as determined by our GALAH DR4 control sample.
Lastly, following \citet{Montalban2002} and \citet{Soares-Furtado2021}, the relative lithium abundance within the engulfed planetary companion is  
\begin{equation}
    10^{A(\text{Li})_\text{p}-12} = \frac{N_\text{p,Li}}{N_\text{p,H}},
\end{equation}
where $A\text{(Li)}_{\text{p}} = 3.35 \pm 0.03$\,dex is assumed to be similar to the proto-solar lithium abundance as inferred from chondritic meteorites \citep{Lodders2020}.
Using the numbers and equations above (and a Monte Carlo estimation described later in this section), the mass of hydrogen in the engulfed planetary companion is $N_\text{p,H} = \left( 3.3 \pm 2.1 \right) \times 10^{30}$\,g.
Assuming a proto-solar hydrogen mass composition of $X_0 = 0.7061$ \citep{Lodders2020}, it is possible to estimate a mass of the engulfed planet as
\begin{equation}
M_{\text{p}} = \frac{m_{\text{H}} N_{\text{p,H}}}{X_{0}}.% = 6.7 \pm 5.0 \, M_{\text{J}}.
\end{equation}

To account for uncertainties, we perform a Monte Carlo analyses with $3\times 10^{5}$ trials. This same number of trials is used for all other Monte Carlo estimates in this paper. For each trial, the convective-zone mass and lithium abundance values (baseline, post-engulfment, and planetary contribution) were drawn from Gaussian distributions.
Our Monte Carlo analysis yielded a median engulfed mass of $5.6^{+5.5}_{-3.5}\,M_{\rm J}$ (16–84\% interval). However, this calculation rests on a critical assumption about planetary composition that we examine in the next section.

\subsubsection{Realistic Heavy-Element Assumption}
\label{subsubsec:realistic}
The proto-solar calculation presented in Section~\ref{subsubsec:protosolar} assumes that planets have proto-solar lithium abundances. 
As acknowledged in \citet{Aguilera_Gomez_2016}, this assumption is inconsistent with both planet formation theory and Solar System observations.
All the solar system planets possess significant metallicity enhancements over a proto-solar composition \citep{Palme2017,Miguel2023}.
This is likely a consequence of the dominant core accretion theory of planet formation, and so an expected outcome for planets across the galaxy \citep{Raymond2022}.
Therefore, independent of stellar lithium depletion processes, this implies that planets are generally enriched in lithium compared to their host stars.
Indeed, terrestrial planets are made effectively entirely of metals \citep[e.g., an enrichment over solar of metals relative to hydrogen in excess of 10$^4$;][]{Marty2012,Halliday2013,Dauphas2013}, while even the giant planets are enriched in heavy elements relative to the Sun by factors between about 5--15 \citep{Miguel2023}.
For simplicity, we could try to account for these considerations with the incorporation of an enrichment factor $\eta$ in the equations above, so that $10^{\text{A(Li)}_\text{p}-12} = \eta N_\text{p,Li} / N_\text{p,H}$, however this approach is fraught since $\eta$ is nonlinearly sensitive to the mass and type of planet and it is not a regularly determined parameter in planetary science.

An alternative approach starts by defining the interacting mass of the engulfed companion $M_i = f_i M_\mathrm{p}$ as the fraction $f_i$ of the engulfed companion of mass $M_\mathrm{p}$ that is incorporated into the convective zone.
Furthermore, we can assert without loss of generality that the mass of engulfed lithium $M_{i,\text{Li}}$ is related to the mass of lithium within the entire engulfed companion $M_{p,\text{Li}}$ according to: 
\begin{equation}
    M_{i,\text{Li}} = f_{i,\text{Li}} M_{p,\text{Li}} \lesssim f_i M_{p,\text{Li}}
\end{equation}
where $f_{i,\text{Li}}$ is the fraction of lithium in the engulfed companion that is incorporated into the convective zone of the star.
To first order, we would expect $f_{i,\text{Li}} \sim f_i$, however if lithium is stored preferentially in one part of a planet, then we might expect that $f_{i,\text{Li}} < f_i$.
Estimating these fractions $f_i$ and $f_{i,\text{Li}}$ is left for future work.

Within the planet, we can relate the lithium abundance to the overall heavy element abundance using relative abundance estimates from cosmochemistry.
The engulfed companion can be divided by mass into hydrogen $X_p$, helium $Y_p$, and heavy (metal) element $Z_p$ mass fractions, so that $X_p +Y_p + Z_p = 1$, and unlike the enrichment factor $\eta$ introduced above, the heavy (metal) element $Z_p$ mass fraction is commonly estimated for different planet types and masses \citep{Palme2014,Helled2022}.
For now, we use chondritic relative abundances since these are the best determined, but we would expect these to vary across the galaxy. 
Given chondritic relative abundances, the mass of lithium within a planet is:
\begin{equation}
    M_{p,\text{Li}} = \left[\text{Li}\right]_\text{p} Z_p M_\mathrm{p}
\end{equation} where the planet's mass abundance of lithium can be estimated from the CI chondrite mass abundance of lithium $\left[\text{Li}\right]_\text{p} = \left[\text{Li}\right]_\text{CI} = 1.5 \pm 0.2 $~$\mathrm{\mu}$g~g$^{-1}$ \citep{Palme2014}, which is representative of the mass abundance of lithium found on average in condensed materials in the solar system.
Thus, the number of lithium atoms added to the star's convective zone from the engulfed planet is:
\begin{equation}
    N_\text{p,Li} = \frac{f_i \left[\text{Li}\right]_\text{CI} Z_p M_\mathrm{p}}{m_\text{Li}}
\end{equation}
where $m_\text{Li}$ is the mean atomic mass of lithium.
Lithium has two stable isotopes with masses: $m_{^6\text{Li}} = 0.999 \times 10^{-23}$\,g and  $m_{^7\text{Li}} = 1.165 \times 10^{-23}$\,g.
Presuming a relative composition similar to the solar system: $\left[^6\text{Li}\right]/\left[^7\text{Li}\right] = 0.082$, then $m_\text{Li} = 1.15 \times 10^{-23}$\,g \citep{Seitz2012}.
Given that planets have metallicity enhancements of $10-10^4$ over proto-solar, we assert that the quantity of hydrogen added by the planet to the convective zone is negligible---we will check this assumption later.

Using the equations and values introduced so far, we can arrange to solve for the mass of the engulfed planet in terms of the measured lithium abundance:
\begin{align}
    M_\mathrm{p} &=  \frac{m_\text{Li}}{f_i \left[\text{Li}\right]_\text{CI} Z_p } \left(N_\text{CZ,Li} - N_\text{s,Li} \right)\\
    &= \frac{X_\text{CZ} M_\text{CZ}  m_\text{Li}}{f_i \left[\text{Li}\right]_\text{CI} Z_p m_\text{H}}\left(\frac{  10^{\text{A(Li)}_\text{CZ}} -10^{\text{A(Li)}_\text{s}} }{ 10^{12}} \right) 
\end{align}
Assuming the incorporated fraction of the engulfed companion into the convective zone is complete ($f_i = 1$), and the engulfed companion has a terrestrial planet-like mass fraction of heavy elements ($Z_p = 1$), our Monte Carlo analysis yields a median engulfed mass of 
$9.4^{+10.7}_{-8.9}\,M_{\oplus}$ (16–84\% interval).
This result suggests that the enhanced lithium signature could be produced by the engulfment of a super-Earth– to Neptune-mass companion.
On the other hand, if the engulfed companion has a giant planet-like mass fraction of heavy elements ($Z_p = 0.1 \pm 0.03$), then our Monte Carlo analyses yields a much larger engulfed mass of $M_p=94.6^{+127.3}_{-90.1}\,M_{\oplus}$ (16–84\% interval).
In both cases, the decision to neglect the planet's contribution to the convective zone hydrogen budget is acceptable, since the planet would contribute less than a percent to the hydrogen mass budget of the convective zone.

The assumed lithium abundance for the engulfed planet is the most critical parameter for determining the engulfed companion mass. 
According to an analysis that acknowledges the significant overall metallicity enhancement within planets \citep[e.g.,][]{Aguilera_Gomez_2016}, we find that the required engulfed planet mass to explain the lithium excess observed in TOI-5882 is between 9--95\,$M_{\oplus}$. 
This is significantly less than the 5.6\,\MJ{} engulfed mass needed to explain the same excess if the planet is assumed to possess a proto-solar relative lithium abundance \citep[e.g.,][]{Montalban2002,Soares-Furtado2021}.

The overall metallicity enhancement of planets rests firmly on the core accretion model for planet formation \citep{Raymond2022}, which is an exceptionally well-tested theoretical framework consistent with observations across the galaxy and particularly well-examined in the context of tightly orbiting planets \citep{Chiang2010,Johansen2017,Liu2020}.
Whereas a model assuming that planets reflect the proto-solar Li/H are much more consistent with a gravitational instability (GI-) model for planet formation.
This planet formation mechanism may be consistent with some rare wide-orbit giant planets \citep{Cadman2021,Speedie2024}, but this theoretical framework struggles to make planets except under special conditions \citep{Boss1997,Rice2003,Zhu2012,Deng2021}.
However, the enhanced metallicity model used here for the lithium abundance $\left[\text{Li}\right]_\text{p}$ within the engulfed planet is based on solar system measurements of CI chondrites $\left[\text{Li}\right]_\text{CI}$, so if condensed material in the TOI-5882 planetary system was depleted or enriched in lithium, this could have a significant effect on the estimated engulfed companion mass, as shown in Equation 13.

\subsection{Future Directions} 
In an ideal scenario, a planetary engulfment candidate like TOI-5882 would be compared to isometallic analogs, such as equal mass stars formed from the same natal molecular cloud, where abundance offsets can more unambiguously be ascribed to external contamination. 
For example, studies of comoving twins and wide binaries have revealed refractory‐metal enhancements that are consistent with planet ingestion \citep[e.g.,][]{KronosKrios2018, widebinaries2020, Liu2024-pb, yana2024}.
However, TOI-5882 lacks comoving companions of a similar mass within a 50\,pc radius (Section \ref{subsec:galah}).
Therefore, we are forced to rely on a large control sample derived from field stars with similar stellar properties.
Fortunately, when lithium enrichment is pronounced, the signal is strong enough to stand out clearly, even within a field population \citep{Soares-Furtado2021}.

Our investigation was limited by the number of lithium equivalent width and abundance measurements among potential control stars. 
Stars spend a small fraction of their lifetimes in the subgiant phase of evolution, therefore, they constitute a small fraction of targets within publicly available survey data. 
While residual inter-pipeline systematics between TRES and GALAH cannot be fully excluded, such effects would need to be both large and systematically aligned with lithium to move TOI-5882 out of the extreme tail of the observed distribution.
To address this gap, future efforts should prioritize a dedicated, homogeneous survey of subgiant stars. Using a single instrument and reduction pipeline tailored to subgiants will minimize systematic errors in stellar parameters. For instance, pipelines optimized for main‐sequence isochrone fitting often perform poorly on subgiants. A unified observing and analysis strategy would establish a consistent baseline of Li\,I EWs and abundances across a large sample, enabling us to determine the true distribution of lithium among subgiants and identify outliers more confidently.

Beyond lithium, examining other elemental tracers would help to identify the enrichment mechanism (See Section \ref{sec:intro} and \citealt{sayeed2023roadsleadlithiumformation}). Our team noted deeper absorption features in the iron-peak refractory elements Ti, V, Mn, and Ni lines in the spectra of TOI-5882 when compared to control stars. 
However, we emphasize that this is a qualitative observation and is not used as evidence in this work. The expected magnitude of abundance changes in elements other than lithium is generally modest for near-solar composition stars and planets. 
Recent theoretical work shows that while lithium is the most sensitive tracer of engulfment, detectable changes in certain refractory elements may occur under favorable conditions, particularly for stars with relatively thin convective envelopes \citep{Lane2026}.

While a full investigation of these iron-peak features lies outside the scope of the present paper, we plan to undertake a line-by-line differential analysis in future work. By measuring the same transitions in TOI-5882 and in each control star, we can cancel out systematic uncertainties in oscillator strengths, continuum placement, and model atmospheres, thereby confirming whether TOI-5882 truly exhibits iron-peak overabundances. A larger control sample will improve statistical significance and help us understand whether these signatures arise from planetary engulfment, stellar evolution effects, or local chemical inhomogeneities.

\section{Summary}\label{sec:summary}

We tested the planetary engulfment hypothesis for TOI-5882, a system selected because it occupies the theoretically favorable evolutionary window to detect lithium enrichment identified by \citet{Soares-Furtado2021} and hosts a massive brown dwarf companion ($22.01^{+0.61}_{-0.72}$\,\MJ{}, $P = 7.1$\,d) capable of driving planetary migration. 
Using TRES spectroscopy, we measured a Li I equivalent width of $75.39 \pm 3.58$\,m\AA{} and abundance of $A(\mathrm{Li}) = 2.49 \pm 0.12$\,dex. 
Comparing the target to \populationsize{} GALAH DR4 control stars with similar stellar properties, we find it ranks in the \percentile{} percentile in both metrics. Tail probability analysis confirms this enhancement is statistically rare.
Among the control population, only one star exhibits higher lithium, resulting in an empirical tail probability of $P = 0.016$ (1.6\%). Bootstrap resampling establishes a median percentile rank (with a 95\% confidence interval) of $99.09^{+0.91}_{-3.64}$\%.
Even at the conservative lower bound, TOI-5882's lithium level would be exhibited by fewer than 5\% of comparable subgiants.

Our MESA stellar models of TOI-5882 indicate a convective zone mass and depth capable of tidally disrupting companions of $\rho_p \lesssim 5\rho_\star$ above the stellar surface or within the convective zone, thereby depositing material where it can be detected in the stellar photosphere. Planet mass-radius relations indicate that such planets would occupy a mass range of $7\,M_\oplus \lesssim M_p \lesssim 1\,M_{\mathrm{J}}$.

Our analysis demonstrates that the mass of a planetary companion required to reproduce the observed lithium enhancement in TOI-5882 is strongly dependent on the assumed planetary lithium abundance. 
Assuming a proto-solar composition, results in an estimated engulfed mass of $\sim 5.6\,M_{\rm J}$. 
However,  this assumption is inconsistent with planet formation theory and Solar System observations. 
We present a framework incorporating realistic planetary compositions by accounting for heavy-element enrichment and using CI chondritic lithium abundances. 
For terrestrial composition ($Z_p = 1$), we estimate a required engulfed mass of $9.4^{+10.7}_{-8.9}\,M_{\oplus}$. For giant planet composition ($Z_p = 0.1$), we estimate a required engulfed mass of $94.6^{+127.3}_{-90.1}\,M_{\oplus}$.
Both estimates fall within the $7\,M_\oplus \lesssim M_p \lesssim 1\,M_{\mathrm{J}}$ range where tidal disruption deposits material in the convective zone, demonstrating physical consistency. 
This result highlights the importance of planetary composition assumptions when interpreting lithium enrichment as an engulfment signature.

While the lithium excess alone is suggestive of an engulfment event, confirming this scenario requires additional lines of evidence.
We suggest future work incorporating a larger spectral survey of subgiant stars, enabling a more robust investigation of stars in this short-lived phase of evolution. 
Further, we suggest analyses of additional chemical tracers, extending beyond lithium; these would provide much-needed constraints on the progenitor pathways and further characterize differences in the chemical distributions of lithium-enriched subgiants. In particular, a detailed, line-by-line differential analysis of refractory-element abundances (e.g., C/N, iron-peak ratios, and condensation-temperature trends) would help distinguish between planetary accretion and alternative enrichment pathways (nova ejecta, cosmic-ray spallation, etc.).
Likewise, measurements of the host's spin-orbit alignment (e.g., via Doppler tomography or Rossiter–McLaughlin) and continued radial-velocity monitoring for outer companions will constrain the brown dwarf's migration history.
TOI-5882 highlights how spectroscopic investigations of systems with massive, close-in companions can reveal fossil records of past dynamical instabilities---insights that are crucial for understanding the ultimate fate of planetary architectures.

\begin{acknowledgments}

We thank J.~Becker, E.~Pass, A.~Stephan, and J.~Steckloff for the insightful conversations related to this work.
We thank the anonymous reviewers for their careful reading and insightful suggestions that greatly improved this manuscript.
This work was catalyzed at the Lamat Institute \citep{2025NatAs...9.1770Q}, an REU program supported by National Science Foundation grant 2150255.
This material is based upon work where BK was supported by the National Science Foundation Graduate Research Fellowship under Grant No.~DGE 2241144 and the Wisconsin Space Grant Consortium under NASA Award No.~80NSSC20M0123. 
MSF gratefully acknowledges the generous support provided by NASA through Hubble Fellowship grant HST-HF2-51493.001-A awarded by the Space Telescope Science Institute, which is operated by the Association of Universities for Research in Astronomy, Inc., for NASA, under the contract NAS 5-26555. 
CAG acknowledges support from Agencia Nacional de Investigación y Desarrollo (ANID), Proyecto FONDECYT Iniciación 11230741.
Support for this research was provided by the Office of the Vice Chancellor for Research and Graduate Education at the University of Wisconsin--Madison with funding from the Wisconsin Alumni Research Foundation.
A portion of this work was performed at the Aspen Center for Physics, which is supported by National Science Foundation grant PHY-2210452.

This research has made use of the Exoplanet Follow-up Observation Program (ExoFOP; DOI: 10.26134/ExoFOP5) website, which is operated by the California Institute of Technology, under contract with the National Aeronautics and Space Administration under the Exoplanet Exploration Program.
We acknowledge the use of public TOI Release data from pipelines at the TESS Science Office and at the TESS Science Processing Operations Center. 
This research has made use of the VizieR catalogue access tool, CDS, Strasbourg, France. The original description of the VizieR service was published in A\&AS 143, 23. 

This work made use of the Fourth Data Release of the GALAH Survey \citep{Buder2021}. The GALAH Survey is based on data acquired through the Australian Astronomical Observatory, under programs: A/2013B/13 (The GALAH pilot survey); A/2014A/25, A/2015A/19, A2017A/18 (The GALAH survey phase 1); A2018A/18 (Open clusters with HERMES); A2019A/1 (Hierarchical star formation in Ori OB1); A2019A/15, A/2020B/23, R/2022B/5, R/2023A/4, R2023B/5 (The GALAH survey phase 2); A/2015B/19, A/2016A/22, A/2016B/10, A/2017B/16, A/2018B/15 (The HERMES-TESS program); A/2015A/3, A/2015B/1, A/2015B/19, A/2016A/22, A/2016B/12, A/2017A/14, A/2020B/14 (The HERMES K2-follow-up program); R/2022B/02 and A/2023A/09 (Combining asteroseismology and spectroscopy in K2); A/2023A/8 (Resolving the chemical fingerprints of Milky Way mergers); and A/2023B/4 (s-process variations in southern globular clusters). We acknowledge the traditional owners of the land on which the AAT stands, the Gamilaraay people, and pay our respects to elders past and present. This paper includes data that has been provided by AAO Data Central (datacentral.org.au).
This work has made use of data from the European Space Agency (ESA) mission \emph{Gaia},\footnote{\url{https://www.cosmos.esa.int/gaia}} processed by the \emph{Gaia} Data Processing and Analysis Consortium (DPAC).\footnote{\url{https://www.cosmos.esa.int/web/gaia/dpac/consortium}} 
\end{acknowledgments}

\vspace{5mm}
\facilities{Fred Lawrence Whipple Observatory (FLWO) 1.5\,m (TRES: Tillinghast Reflector Echelle Spectrograph), 
Gaia DR3 \citep{GaiaDR3}, 
GALAH DR4 \citep{2024arXiv240919858B}, Mikulski Archive for Space Telescopes \citep{MAST}}

\software{
\texttt{astropy} \citep{Astropy:2013, Astropy:2018}, 
\texttt{astroquery} \citep{astroquery}, 
\exofast{} \citep{Eastman2019}, 
\texttt{matplotlib} \citep{matplotlib}, 
\texttt{MESA} \citep{Dotter_2016, Paxton:2011, Paxton:2013, Paxton:2015},
\texttt{MIST} \citep{Choi_2016},
\texttt{numpy} \citep{Harris:2020}, 
\texttt{pandas} \citep{pandas:2023}, 
\texttt{scipy} \citep{Virtanen:2020}, 
\texttt{specutils} \citep{specutils}}

\bibliographystyle{aasjournalv7}
\bibliography{bibliography.bib}

\appendix
\section{Control Sample Spectral Validation}
\label{appendix_spectra}
\begin{figure}[h!]
    \centering
    \includegraphics[width=\linewidth]{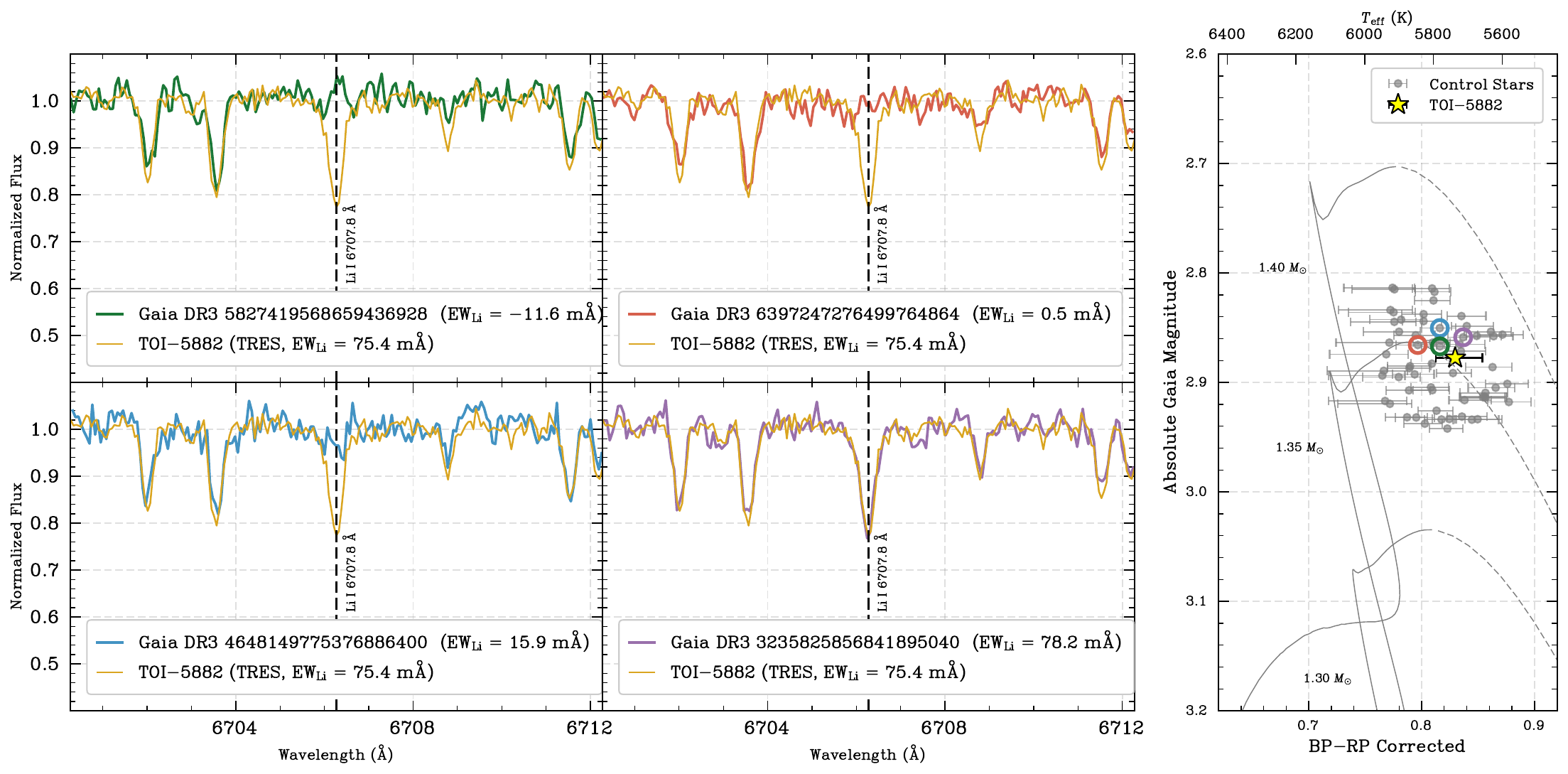}
    \caption{Left: Comparison of the Li~I 6707.8\,\AA\ region for TOI-5882 (gold) and four representative control stars (colored), spanning a range of Li\,I EW  values. 
    The enhanced depth of the Li feature in TOI-5882 is directly evident in the normalized spectra. 
    The control star with the strongest Li absorption provides a close match to the line depth, with Li\,I EW  consistent within measurement uncertainties, while the remaining controls show systematically weaker features. 
    Right: Color--magnitude diagram illustrating the location of TOI-5882 (star symbol) and the selected control stars (circled points) relative to the full comparison sample (gray).}
    \label{fig:spectral_comparison}
\end{figure}

\section{\textsc{MESA} Inlist Parameters}\label{appendix_inlist}
All unspecified parameters are set to \textsc{MESA} defaults (version 24.08.1).
\begin{verbatim}
&star_job
   load_saved_model = .true.
   load_model_filename = `start.mod' !ZAMS star
/ ! end of star_job namelist

&kap
   Zbase = 0.03
   use_Type2_opacities = .true.
/ !end of kap namelist

&controls
   initial_mass = 1.334
   initial_z = 0.03
   solver_iters_timestep_limit = 3
   max_abs_rel_run_E_err = 1d-2
   mixing_length_alpha = 1.73 !we also created models set to 1.65 and 1.8
   varcontrol_target = 1.0d-4
   overshoot_scheme(1) = `exponential' 
   overshoot_zone_type(1) = `any'
   overshoot_zone_loc(1) = `any'
   overshoot_bdy_loc(1) = `any'
   overshoot_f(1) = 0.014
   overshoot_f0(1) = 0.004
   atm_option = `T_tau'
   atm_T_tau_relation = `Eddington'
   atm_T_tau_opacity = `fixed'
   cool_wind_full_on_T = 9.99d9
   hot_wind_full_on_T = 1d10
   cool_wind_RGB_scheme = `Reimers'
   cool_wind_AGB_scheme = `Blocker'
   RGB_to_AGB_wind_switch = 1d-4
   Reimers_scaling_factor = 0.5d0
   Blocker_scaling_factor = 0.0003d0
/ ! end of controls namelist
\end{verbatim}

\end{document}